\documentclass[aps,prd,reprint,nofootinbib,superscriptaddress]{revtex4-2}

\usepackage{amsmath,amssymb,bm}
\usepackage{graphicx}
\usepackage{hyperref}
\usepackage{xcolor}
\usepackage{booktabs}
\usepackage{mathtools}

\newcommand{\dd}{\mathrm{d}}
\newcommand{\ee}{\mathrm{e}}
\newcommand{\mcR}{\mathcal{R}}

\begin{document}

\title{Bounds on the Lyapunov Exponent of Circular Null Orbits in $n$-Dimensional Black-Hole Spacetimes}

\author{Anagha V.}
\email{anagha.v.physics@gmail.com}
\affiliation{Department of Physics, T. K. M. College of Arts and Science, Kollam, Kerala, India}
\author{C. Fairoos}
\email{fairoos.phy@gmail.com}
\affiliation{Department of Physics, T. K. M. College of Arts and Science, Kollam, Kerala, India}
\author{T. K. Safir}
\email{stkphy@gmail.com}
\affiliation{Department of Physics, T. K. M. College of Arts and Science, Kollam, Kerala, India}

%\date{\today}

\begin{abstract}
Unstable circular null orbits provide a geometric bridge between black-hole optics, photon rings, strong gravitational lensing, and the eikonal sector of quasinormal ringing.  In four-dimensional Einstein gravity, the instability rate of such orbits, measured by the Lyapunov exponent $\lambda$, obeys model-independent upper bounds when the matter sector satisfies standard energy conditions.  We extend this analysis to static, spherically symmetric, asymptotically flat black holes in arbitrary dimensional Einstein gravity, allowing for an anisotropic matter distribution.  We show that, under the tangential null energy condition, the Lyapunov exponent admits a dimension-dependent upper bound in terms of the generalized surface gravity $\kappa (r)$ and the metric function $\mu(r)$, both evaluated at the photon sphere $r =r_\gamma$. For $n$- dimensional black hole spacetime, the bound is $\lambda\leq\sqrt{n-3}\, {\kappa_\gamma}/{\sqrt{\mu_\gamma}}$. Related bounds are derived in terms of the critical impact parameter (shadow radius), the orbital frequency, a local acceleration scale, and eikonal quasinormal-mode damping.  All principal inequalities reduce to the known four-dimensional results for $n=4$. Also, the bounds involving the critical impact parameter and orbital frequency are saturated by the Schwarzschild-Tangherlini vacuum solution.  The results provide a compact set of consistency conditions linking the dimensionality of spacetime to the instability of photon trapping in Einstein gravity.
\end{abstract}

\maketitle

\section{Introduction}
\label{sec:intro}

Circular null geodesics are among the simplest structures that encode genuinely strong-field information about a black-hole spacetime \cite{Chandrasekhar1983}. In a static, spherically symmetric geometry, the unstable circular null orbit defines the photon sphere and determines the critical impact parameter separating capture from escape. The unstable null orbit also enters several observable properties of black holes. It determines the critical impact parameter associated with the shadow, governs the strong-deflection behavior of gravitational lensing \cite{Bozza2002,Stefanov2010}, and plays an important role in the eikonal description of black-hole wave dynamics \cite{Cardoso2009,Decanini2011}.\\

%The subject has acquired additional relevance because black-hole strong-field observables are no longer only theoretical constructs.  The Event Horizon Telescope (EHT) has resolved horizon-scale emission around M87$^*$ and Sgr A$^*$, providing measurements of ring-like structures whose characteristic angular scales are tied to the critical photon region \cite{EHTM87I2019,EHTM87VI2019,EHTSgrA2022}.  At the same time, gravitational-wave observations of binary black-hole mergers have made the ringdown regime accessible to experiment, and the observed post-merger signal is consistent with black-hole quasinormal ringing within present uncertainties \cite{LIGOTests2016,BertiCardosoStarinets2009}.  Future very-long-baseline interferometry and space-based gravitational-wave observations are expected to sharpen both photon-ring and ringdown tests \cite{Johnson2020,Walia2026}.  These developments make model-independent relations among null-orbit geometry, instability rates, shadows, and quasinormal modes especially valuable.

Recent observations provide additional motivation for studying the properties of unstable null orbits. The Event Horizon Telescope has resolved horizon-scale emission around M87$^*$ and Sgr A$^*$, with ring-like structures probing the strong-field region close to the black hole \cite{EHTM87I2019,EHTM87VI2019,EHTSgrA2022}. Gravitational-wave 
observations of binary black-hole mergers have likewise opened the ringdown regime to observational tests, with the post-merger signal being consistent with black-hole quasinormal ringing within current uncertainties \cite{LIGOTests2016,BertiCardosoStarinets2009}. These observations motivate relations that connect the geometry and instability of null orbits with quantities associated with black-hole imaging and ringdown.\\

%
%The instability of a circular null geodesic can be characterized by a Lyapunov exponent $\lambda$.  A positive $\lambda$ measures the exponential separation of neighboring radial trajectories and therefore sets an inverse instability timescale \cite{Cardoso2009}.  In the eikonal regime, the same exponent determines the imaginary part of the quasinormal-mode frequency, whereas the real part is controlled by the angular frequency of the circular null orbit \cite{Cardoso2009}.  This relation provides a direct physical interpretation of the photon-sphere instability: a larger $\lambda$ corresponds to more rapid leakage of radiation from the photon region and, within the geodesic/WKB correspondence, faster damping of the associated modes.

The instability of a circular null geodesic is characterized by its Lyapunov exponent $\lambda$. A positive $\lambda$ describes the exponential separation of nearby radial trajectories and defines the instability timescale $T_\lambda=1/\lambda$ \cite{Cardoso2009}. In the eikonal regime, the real and imaginary parts of the quasinormal-mode frequency are related, respectively, to the orbital frequency and the Lyapunov exponent of the unstable null orbit \cite{Cardoso2009}. Thus, whenever the standard geodesic/QNM correspondence applies, the same quantity that measures null-orbit instability also determines the damping rate of the corresponding eikonal modes.\\

%Bounds on Lyapunov exponents are also conceptually interesting in the 
%broader context of gravitational chaos. The celebrated quantum-chaos 
%inequality of Maldacena, Shenker, and Stanford constrains thermal 
%many-body systems by $\lambda\leq 2\pi T$ under appropriate assumptions 
%\cite{Maldacena2016}, while near-horizon particle dynamics has motivated 
%related gravitational bounds involving the horizon surface gravity 
%\cite{Hashimoto2017}. Although the Lyapunov exponent entering the 
%many-body chaos bound is conceptually distinct from the geodesic 
%Lyapunov exponent considered here, these results motivate the search 
%for universal restrictions on instability in gravitational systems. 
%Circular null orbits, however, need not obey a simple bound in terms 
%of the horizon surface gravity. Recently, Gallo and M\"adler established 
%alternative universal inequalities for static, spherically symmetric 
%black holes in four-dimensional Einstein gravity, relating the 
%photon-sphere Lyapunov exponent to a generalized surface gravity, a 
%local acceleration scale, and the shadow radius 
%\cite{GalloMadler2025}.

The possibility of bounding instability rates has appeared in several contexts. Maldacena, Shenker, and Stanford obtained the bound $\lambda\leq2\pi T$ for the Lyapunov exponent characterizing quantum chaos in thermal many-body systems under appropriate assumptions \cite{Maldacena2016}. Related studies of particle motion near black-hole horizons have led to bounds involving the horizon surface gravity \cite{Hashimoto2017}. The Lyapunov exponent considered in the present work is instead associated with the radial instability of a circular null geodesic, and such orbits do not in general satisfy a bound set by the horizon surface gravity. Gallo and M\"adler recently showed that, for static and spherically symmetric black holes in four-dimensional Einstein gravity satisfying suitable energy conditions, $\lambda$ obeys a different set of bounds involving a generalized surface gravity at the photon sphere, the proper acceleration of static observers, and the shadow radius \cite{GalloMadler2025}.\\

%The question of how such bounds extend to higher dimensions has also 
%received attention. In particular, Bianchi, Grillo, and Morales studied 
%the instability of massless trajectories near photon spheres in 
%higher-dimensional black-hole and fuzzball geometries and obtained a 
%dimension-dependent upper bound on the Lyapunov exponent in terms of 
%the critical impact parameter \cite{Bianchi:2020des}. Their 
%analysis, however, was carried out for specific classes of geometries. 
%A corresponding bound for a general static, spherically symmetric 
%black hole in higher-dimensional Einstein gravity, allowing for an 
%anisotropic matter distribution, has not been established. This provides 
%the motivation for the present work.

Higher-dimensional Lyapunov bounds have also been considered for specific black-hole geometries. Bianchi, Grillo, and Morales studied massless-particle dynamics near photon spheres of several higher-dimensional black-hole and fuzzball solutions and obtained a dimension-dependent upper bound involving the critical impact parameter \cite{Bianchi:2020des}. Their analysis indicates that the dimensionality of spacetime enters explicitly in the maximal instability of circular null motion. Here we address the problem in a different setting: rather than specifying a particular black-hole solution, we consider a general static, spherically symmetric solution of the $n$-dimensional Einstein equations sourced by anisotropic matter and derive the bound directly from the field equations and energy conditions.\\

%Higher-dimensional black holes arise naturally in string-inspired 
%models, brane-world scenarios, Kaluza--Klein constructions, and 
%gauge/gravity duality, and also provide a useful setting in which to 
%test which properties of four-dimensional black holes persist when the 
%spacetime dimension is changed \cite{Tangherlini1963,EmparanReall2008}. 
%Photon spheres and their radius bounds have already been investigated 
%in higher-dimensional Einstein and Einstein--Gauss--Bonnet gravity 
%\cite{GalloVillanueva2015}. More recently, Song, Fu, and Cen derived 
%dimension-dependent upper and lower bounds on photon-sphere radii for 
%$n$-dimensional Einstein black holes under suitable assumptions on the 
%matter fields \cite{SongFuCen2026}. These developments provide the 
%geometrical framework for investigating whether the four-dimensional 
%Lyapunov bounds admit a similarly general higher-dimensional extension.

The extension to higher dimensions is particularly relevant here because photon-sphere properties themselves acquire an explicit dependence on the spacetime dimension. Gallo and Villanueva studied photon spheres in $n$-dimensional Einstein and Einstein-Gauss-Bonnet gravity \cite{GalloVillanueva2015}, while Song, Fu, and Cen recently obtained dimension-dependent upper and lower bounds on the photon-sphere radius for static, spherically symmetric black holes in $n$-dimensional Einstein gravity under appropriate assumptions on the matter fields \cite{SongFuCen2026}. Higher-dimensional black holes also arise in several extensions and applications of general relativity, including Kaluza-Klein, brane-world, 
and string-inspired settings \cite{EmparanReall2008}. It is therefore natural to examine how the four-dimensional Lyapunov bounds are modified in the same general higher-dimensional setting.\\

The main result of this work is the bound
\begin{equation}
 \lambda\leq \sqrt{n-3}\,
 \frac{\kappa_\gamma}{\sqrt{\mu_\gamma}},
 \label{eq:mainbound}
\end{equation}
for an unstable circular null orbit at $r=r_\gamma$. Although the class of matter configurations considered below satisfies the dominant energy condition, the step leading to this inequality requires only the local condition $(\rho+p_t)_{r_\gamma}\geq0$. Here $\mu_\gamma=\mu(r_\gamma)$ and $\kappa_\gamma$ denotes the generalized surface gravity evaluated at the photon sphere. The same bound can be expressed in terms of the critical impact parameter and the orbital frequency as
\begin{equation}
 \lambda R_{\rm sh}\le\sqrt{n-3},
 \qquad
 \frac{\lambda}{\Omega_\gamma}\le\sqrt{n-3},
 \label{eq:introshadow}
\end{equation}
where $R_{\rm sh}$ denotes the critical impact parameter, which determines the shadow radius for an observer at infinity, and $\Omega_\gamma$ is the coordinate angular frequency of the circular null orbit.  The Schwarzschild-Tangherlini solution saturates these normalized bounds, showing that the factor $\sqrt{n-3}$ is sharp within the class considered.\\

%The paper is organized as follows.  In Sec.~\ref{sec:setup} we introduce the $n$-dimensional Einstein system and derive the photon-sphere condition following Ref.~\cite{SongFuCen2026}.  Section~\ref{sec:lyapunov} reviews the dynamical meaning of the Lyapunov exponent and obtains its null-orbit expression following Refs.~\cite{Cardoso2009,GalloMadler2025,GalloVillanueva2015}.  In Sec.~\ref{sec:bound} we derive the dimension-dependent upper bound and its generalized-surface-gravity form.  Section~\ref{sec:applications} develops additional consequences for the shadow radius, orbital frequency, critical exponent, local acceleration scale, and eikonal quasinormal modes.  Section~\ref{sec:checks} discusses the four-dimensional limit, Schwarzschild--Tangherlini saturation, and stronger corollaries that follow under additional matter assumptions.  We present our conclusions in Sec.~\ref{sec:conclusions}.  Natural units $G=c=\hbar=1$ are used throughout.

The paper is organized as follows. Section~\ref{sec:setup} introduces the $n$-dimensional Einstein system and the photon-sphere condition. In Sec.~\ref{sec:lyapunov} we review the Lyapunov description of circular null-orbit instability, and in Sec.~\ref{sec:bound} we derive the dimension-dependent upper bound. Its consequences for the shadow radius, orbital frequency, local acceleration, critical exponent, and eikonal quasinormal modes are discussed in Sec.~\ref{sec:applications}. Consistency checks and limiting cases are presented in Sec.~\ref{sec:checks}, followed by the conclusions in Sec.~\ref{sec:conclusions}. We use units $G=c=\hbar=1$.\\

\section{Higher-dimensional geometry and the photon sphere}
\label{sec:setup}

\subsection{Metric and field equations}

We consider a static, spherically symmetric, asymptotically flat black-hole spacetime in $n\ge4$ dimensions, described by \cite{SongFuCen2026,GalloVillanueva2015}
\begin{equation}
 \dd s^2=-\ee^{-2\delta(r)}\mu(r)\dd t^2+\frac{\dd r^2}{\mu(r)}+r^2\dd\Omega_{n-2}^2,
 \label{eq:metric}
\end{equation}
with $\dd\Omega_{n-2}^2$ being the metric of the $(n-2)$-sphere. Note that the function $\delta(r)$ is retained in order to accommodate a general static, spherically symmetric matter distribution. The event horizon $r=r_H$ is characterized by $\mu(r_H)=0$. Also, we require the following conditions at the event horizon \cite{Nunez:1996xv}:
\begin{equation}
\mu'(r_H) \geq0, \quad \delta(r_H) \leq \infty \quad \delta'(r_H) \leq \infty
\end{equation}

Asymptotic flatness requires,
\begin{equation}
    \mu(r)\to1,
    \qquad
    \delta(r)\to0
    \qquad
    \text{as } r\to\infty.
     \label{eq:asymptotic}
\end{equation}

The gravitational field equations in $n$-dimensional Einstein gravity are
\begin{equation}
    G_{\mu\nu}=8\pi T_{\mu\nu},
    \label{eq:einstein}
\end{equation}
where $G_{\mu\nu}$ is the Einstein tensor and $T_{\mu\nu}$ represents 
the energy-momentum tensor of the matter fields. We model the matter 
distribution as an anisotropic fluid, for which the energy-momentum 
tensor takes the form
\begin{equation}
    T^{\mu}{}_{\nu}
    =
    \mathrm{diag}
    \left(
    -\rho(r),\,p_r(r),\,
    \underbrace{p_t(r),\ldots,p_t(r)}_{n-2}
    \right).
    \label{eq:stress_tensor}
\end{equation}
Here, $\rho(r)$ is the energy density, while $p_r(r)$ and $p_t(r)$ 
denote the radial and tangential pressures, respectively. The 
tangential pressure is the same along each of the $(n-2)$ angular 
directions. In general, $p_r$ and $p_t$ need not be equal, allowing 
the matter distribution to be anisotropic. The trace of the energy-momentum tensor 
in $n$ dimensions follows directly from Eq.~\eqref{eq:stress_tensor} 
and is given by
\begin{equation}
    T\equiv T^{\mu}{}_{\mu}
    =
    -\rho+p_r+(n-2)p_t.
    \label{eq:trace}
\end{equation}

For the metric introduced above, the $tt$ and $rr$ components of 
Eq.~\eqref{eq:einstein} yield
\begin{equation}
    \mu'
    =
    \frac{(n-3)(1-\mu)}{r}
    -
    \frac{16\pi r}{n-2}\rho ,
    \label{eq:mup}
\end{equation}
and
\begin{equation}
    \delta'
    =
    -\frac{8\pi r}{(n-2)\mu}
    \left(\rho+p_r\right).
    \label{eq:deltap}
\end{equation}
These equations relate the radial variation of the metric functions to the density and pressure of the matter surrounding the black hole.The function $\mu$ relates to the generalized Misner-Sharp mass $m(r)$ through \cite{MisnerSharp1964, Zhang:2014goa}
\begin{equation}
    \mu(r)
    =
    1-\frac{2m(r)}{r^{n-3}},
    \label{eq:mass_metric}
\end{equation}
where,
\begin{equation}
    m(r)
    =
    m_H
    +
    \frac{8\pi}{n-2}
    \int_{r_H}^{r}x^{n-2}\rho(x)\,dx .
    \label{eq:mass_function}
\end{equation}
Here, $m_H$ denotes the horizon mass. The function $m(r)$ therefore measures the mass contained within a sphere of radius $r$, including the contribution of the matter distribution outside the event horizon. The ADM mass $M$ is obtained from the asymptotic value of the mass function,
\begin{equation}
    M=\lim_{r\rightarrow\infty}m(r).
\end{equation}
The finiteness of this limit requires the energy density to fall off sufficiently rapidly at spatial infinity. In particular,
\begin{equation}
    \lim_{r\rightarrow\infty} r^{\,n-1}\rho(r)=0.
\end{equation}

We require the matter fields to satisfy the dominant energy condition
\cite{HawkingEllis,Wald},
\begin{equation}
    \rho\geq0,
    \qquad
    \rho\geq |p_r|,
    \qquad
    \rho\geq |p_t|.
    \label{eq:DEC}
\end{equation}
At the event horizon, regularity imposes an additional relation between
the energy density and radial pressure. Evaluating the corresponding
Einstein equations at $r=r_H$, where $\mu(r_H)=0$, gives
\begin{equation}
    \rho(r_H)+p_r(r_H)=0.
    \label{eq:horizon_pressure}
\end{equation}
Thus, the radial pressure at the horizon is fixed by the local energy
density, $p_r(r_H)=-\rho(r_H)$.\\

In what follows, we briefly review the condition for circular null orbits in an $n$-dimensional black-hole spacetime.

\subsection{Null geodesics and the circular-orbit condition}

We now turn to the motion of massless particles and derive the condition for the existence of circular null orbits. Owing to the spherical symmetry of the spacetime, the orbital plane can always be chosen such that the motion is confined to an equatorial plane. This choice involves no loss of generality and considerably simplifies the geodesic equations. We therefore fix the angular coordinates according to
\begin{equation}
    \theta_1=\theta_2=\cdots=\theta_{n-3}
    =\frac{\pi}{2},
    \qquad
    \theta_{n-2}=\phi,
    \label{eq:equatorial_plane}
\end{equation}
where $\phi$ denotes the azimuthal coordinate. With this choice, the 
motion of a test particle is described entirely by the coordinates 
$t$, $r$, and $\phi$. The Lagrangian governing null geodesics in the 
spacetime introduced above can then be written as
\begin{equation}
 2\mathcal L=-\ee^{-2\delta}\mu\dot t^{\,2}+\frac{\dot r^{\,2}}{\mu}+r^2\dot\phi^{\,2}=0,
 \label{eq:nullLagrangian}
\end{equation}
where the dot denotes differentiation with respect to an affine parameter.  The two Killing symmetries imply the conserved quantities
\begin{equation}
 E=\ee^{-2\delta}\mu\dot t,
 \qquad
 L=r^2\dot\phi.
 \label{eq:EL}
\end{equation}
Using these in Eq.~\eqref{eq:nullLagrangian}, the radial equation may be written as
\begin{equation}
 \dot r^{\,2}=V(r)
 =\mu\left(\frac{E^2}{\ee^{-2\delta}\mu}-\frac{L^2}{r^2}\right).
 \label{eq:radialpotential}
\end{equation}
A circular null orbit satisfies \cite{osti_4585183},
\begin{equation}
 V(r_\gamma)=0,
 \qquad
 V'(r_\gamma)=0.
 \label{eq:circularconditions}
\end{equation}
These conditions yield \cite{SongFuCen2026}
\begin{equation}
 -r\mu'+2\mu(1+r\delta')=0.
 \label{eq:geometricPS}
\end{equation}
Equivalently, if
\begin{equation}
 A(r)\equiv\ee^{-2\delta(r)}\mu(r),
 \label{eq:Adef}
\end{equation}
then Eq.~\eqref{eq:geometricPS} becomes
\begin{equation}
 \left.\frac{A'}{A}\right|_{r_\gamma}=\frac{2}{r_\gamma}.
 \label{eq:Aphoton}
\end{equation}
Here, $r_\gamma$ represents the radius of the photon sphere.

Substituting the Einstein equations \eqref{eq:mup} and \eqref{eq:deltap} into Eq.~\eqref{eq:geometricPS}, one obtains the photon-sphere characteristic function introduced in higher-dimensional form in Refs.~\cite{GalloVillanueva2015,SongFuCen2026},
\begin{equation}
  \mcR(r)=3-n+(n-1)\mu(r)-\frac{16\pi r^2p_r(r)}{n-2}
 \label{eq:Rdef}
\end{equation}
for the photon sphere,
\begin{equation}
 \mcR(r_\gamma)=0.
 \label{eq:Rzero}
\end{equation}
The same quantity has the purely geometric representation
\begin{equation}
 \mcR(r)=2\mu-r\mu'+2r\mu\delta'
 =\ee^{2\delta}\left[2A-rA'\right],
 \label{eq:Rgeometric}
\end{equation}

Thus, Eq.~\eqref{eq:Rzero} is simply the circular-null-orbit condition given in Eq.\eqref{eq:Aphoton}, expressed in terms of the characteristic function $\mathcal{R}(r)$.

For a regular outer horizon, one has $\mcR(r_H)\leq0$, whereas 
asymptotic flatness together with a sufficiently rapid falloff of the 
matter fields gives $\mcR(\infty)=2$. Continuity therefore guarantees 
the existence of at least one zero of $\mcR(r)$ outside the event 
horizon. For the class of black-hole spacetimes considered in 
Ref.~\cite{SongFuCen2026}, the location of the photon sphere is further 
constrained by the dimension-dependent bounds
\begin{equation}
    \left(\frac{n-1}{2}\right)^{\frac{1}{n-3}}r_H
    \leq r_\gamma
    \leq [(n-1)M]^{\frac{1}{n-3}},
\end{equation}
where the lower bound requires the additional assumption that 
$|r^{n-1}p_r(r)|$ decreases monotonically outside the horizon. These 
bounds reduce to the familiar four-dimensional results 
$3r_H/2\leq r_\gamma\leq3M$ when $n=4$ \cite{Hod:2013jhd, Hod:2020pim}.
\section{Lyapunov exponent of circular null geodesics}
\label{sec:lyapunov}

\subsection{Linear stability in phase space}

The Lyapunov exponent quantifies the exponential growth or decay of infinitesimal perturbations around a reference trajectory.  Following Cardoso \emph{et al.} \cite{Cardoso2009}, consider a dynamical system
\begin{equation}
 \frac{\dd X_i}{\dd t}=H_i(X_j).
 \label{eq:dynamicalsystem}
\end{equation}
If $X_i(t)$ is an exact trajectory and $X_i(t)+\delta X_i(t)$ is a nearby one, linearization gives
\begin{equation}
 \frac{\dd\,\delta X_i}{\dd t}=K_{ij}(t)\delta X_j,
 \qquad
 K_{ij}=\left.\frac{\partial H_i}{\partial X_j}\right|_{X(t)}.
 \label{eq:linearized}
\end{equation}
A solution to the linearized equation can be written as
\begin{equation}
    \delta X_i(t)=L_{ij}(t)\delta X_j(0),
\end{equation}
where the evolution matrix $L_{ij}(t)$ satisfies
\begin{equation}
    \dot{L}_{ij}(t)=K_{im}L_{mj}(t),
    \qquad
    L_{ij}(0)=\delta_{ij}.
     \label{eq:evolution}
\end{equation}

The principal Lyapunov exponent measures the asymptotic logarithmic growth rate of the perturbation.  For the two-dimensional radial phase space $X=(p_r,r)$ associated with a circular orbit, the linear stability matrix is off diagonal and the nontrivial exponents are $\lambda=\pm\sqrt{K_1K_2}$ \cite{Cardoso2009}.  Expressed through the radial potential $V$ defined by $\dot r^{\,2}=V(r)$, the coordinate-time Lyapunov exponent is
\begin{equation}
 \lambda^2=\left.\frac{V''}{2\dot t^{\,2}}\right|_{r_\gamma}.
 \label{eq:CardosoLambda}
\end{equation}
A real and positive $\lambda$ therefore characterizes the instability of 
the circular null orbit, with the corresponding instability timescale 
given by $T_\lambda=1/\lambda$. Our next goal is to relate this instability 
directly to the geometry of the photon sphere. In particular, we express 
the Lyapunov exponent in terms of the characteristic function $\mcR(r)$ 
introduced in the previous section. This relation will provide the 
starting point for deriving an upper bound on $\lambda$.

\subsection{Expression in terms of the photon-sphere function}

 The second derivative of the radial effective potential can be simplified 
considerably at the circular null orbit. Differentiating the effective 
potential twice with respect to $r$ and making use of the circular-orbit 
conditions, one obtains \cite{GalloVillanueva2015}
\begin{equation}
\begin{split}
V''(r_\gamma)
={}&\frac{L^2}{r_\gamma^4 \ e^{-2\delta(r_\gamma)}}\,
\Bigg\{
e^{-2\delta(r_\gamma)}\mu(r_\gamma)
\\
&\qquad
-r_\gamma^2
\left[
e^{-2\delta(r_\gamma)}\mu(r_\gamma)
\right]''
\Bigg\}.
\end{split}
\label{eq:Vpp}
\end{equation}
Here, the circular-orbit condition has been used to eliminate the terms 
containing the first radial derivative of the metric functions. This 
expression relates the instability of the orbit directly to the local 
radial variation of the metric near the photon sphere.

The result can be brought into a more useful form by making use of the 
characteristic function $\mcR(r)$ introduced in the previous section. 
A straightforward calculation using its definition gives
\begin{equation}
    V''(r_\gamma)
    =
    \frac{L^2}{r_\gamma^3}\,
    \mcR'(r_\gamma).
    \label{eq:VppR}
\end{equation}
Thus, the stability of the circular null orbit is encoded in the radial 
derivative of the characteristic function evaluated at the photon sphere.
Using $V(r_\gamma)=0$ together with $E=A\dot t$ gives
\begin{equation}
 \dot t_\gamma^2=\frac{L^2}{r_\gamma^2A_\gamma},
 \qquad
 A_\gamma=\ee^{-2\delta_\gamma}\mu_\gamma.
 \label{eq:tdotgamma}
\end{equation}
Equations \eqref{eq:CardosoLambda}-\eqref{eq:tdotgamma} then yield
\begin{equation}
 \lambda^2=\left.\frac{\ee^{-2\delta}\mu}{2r}\mcR'(r)\right|_{r=r_\gamma}.
 \label{eq:lambdabase}
\end{equation}

Equation~\eqref{eq:lambdabase} expresses the Lyapunov exponent directly in terms of the radial derivative of the characteristic function, in agreement with the $n$-dimensional result of Ref.~\cite{GalloVillanueva2015}. The problem of bounding $\lambda$ is therefore reduced to controlling $\mathcal{R}'(r_\gamma)$.

\section{A dimension-dependent upper bound}
\label{sec:bound}

For the anisotropic energy-momentum tensor introduced above, the radial 
component of the conservation equation,
\begin{equation}
    \nabla_{\mu}T^{\mu}{}_{r}=0,
\end{equation}
gives \ref{app:pressure_conservation},
\begin{equation}
    p_r'
    =
    -\frac{A'}{2A}\left(\rho+p_r\right)
    +\frac{n-2}{r}\left(p_t-p_r\right).
    \label{eq:conservationA}
\end{equation}
Here, the first term describes the contribution associated with the radial 
variation of the redshift factor, whereas the second term arises from the 
pressure anisotropy in the $(n-2)$ angular directions.  At the photon sphere, we use the circular-null-orbit condition derived in Eq.~\eqref{eq:Aphoton}. Consequently, Eq.~\eqref{eq:conservationA} evaluated at $r=r_\gamma$ becomes

\begin{equation}
 p_r'(r_\gamma)=\frac{-\rho_\gamma+(n-2)p_{t\gamma}-(n-1)p_{r\gamma}}{r_\gamma}.
 \label{eq:prprimegamma}
\end{equation}
Differentiating Eq.~\eqref{eq:Rdef},
\begin{equation}
 \mcR'=(n-1)\mu'
 -\frac{16\pi}{n-2}\left(2rp_r+r^2p_r'\right).
 \label{eq:Rprime1}
\end{equation}
At the circular null orbit, Eq.~\eqref{eq:Rzero} implies
\begin{equation}
 (n-1)\mu_\gamma=(n-3)+\frac{16\pi r_\gamma^2p_{r\gamma}}{n-2}.
 \label{eq:muphotonrelation}
\end{equation}
Substituting Eqs.~\eqref{eq:mup}, \eqref{eq:prprimegamma}, and \eqref{eq:muphotonrelation} into Eq.~\eqref{eq:Rprime1}, all radial-pressure terms cancel.  One obtains the exact identity
\begin{equation}
 \mcR'(r_\gamma)=\frac{2(n-3)}{r_\gamma}
 -16\pi r_\gamma\left(\rho_\gamma+p_{t\gamma}\right).
 \label{eq:RprimeExact}
\end{equation}
This cancellation is the central technical observation of the present analysis. If the tangential null energy condition holds at the photon sphere,
\begin{equation}
 \rho_\gamma+p_{t\gamma}\ge0,
 \label{eq:tangentialNEC}
\end{equation}
then Eq.~\eqref{eq:RprimeExact} immediately gives
\begin{equation}
 \mcR'(r_\gamma)\le\frac{2(n-3)}{r_\gamma}.
 \label{eq:RprimeBound}
\end{equation}
Notice that no separate sign assumption on $p_r(r_\gamma)$ is required for this particular instability bound.  The radial pressure enters the location of the photon sphere through Eq.~\eqref{eq:Rzero}, but it drops out of the local upper bound on $\mcR'(r_\gamma)$.

Now, the bound on Lyapunov exponent can be obtained by substituting Eq.~\eqref{eq:RprimeBound}  into Eq.~\eqref{eq:lambdabase} yielding,
\begin{equation}
 \lambda^2\le(n-3)\frac{\ee^{-2\delta_\gamma}\mu_\gamma}{r_\gamma^2}.
 \label{eq:rawbound}
\end{equation}
Equivalently,
\begin{equation}
 \lambda\le\sqrt{n-3}\,
 \frac{\ee^{-\delta_\gamma}\sqrt{\mu_\gamma}}{r_\gamma}.
 \label{eq:rawboundsqrt}
\end{equation}
This establishes the principal dimension-dependent bound on the Lyapunov exponent. To bring out its physical significance, we now recast this result in terms of quantities characterizing the local geometry and optical properties of the circular null orbit.\\

Following Ref.~\cite{AbreuVisser2010}, we introduce a generalized notion of surface gravity away from the event horizon. The construction is based on the proper acceleration of a static observer. Unlike a freely falling observer, an observer held at a fixed radial position must maintain a nonzero acceleration in order to remain static in the gravitational field. The magnitude of this acceleration provides a natural local measure of the gravitational field strength. After accounting for the gravitational redshift between the observer and infinity, it can be used to define a generalized surface gravity at an arbitrary radius.\\

A static observer at a fixed radial position follows the integral curves of the timelike Killing vector. For the metric \eqref{eq:metric}, with $A(r)=e^{-2\delta(r)}\mu(r)$, the normalized four-velocity is
\begin{equation}
    u^\mu=\left(A^{-1/2},0,\ldots,0\right).
\end{equation}
Although the observer has no spatial coordinate velocity, its worldline is not geodesic. The corresponding four-acceleration, $a^\mu=u^\nu\nabla_\nu u^\mu$, has only a radial component,
\begin{equation}
    a^r
    =
    \Gamma^r{}_{tt}(u^t)^2.
\end{equation}
Using $\Gamma^r{}_{tt}=\mu A'/2$, together with $u^t=A^{-1/2}$, one obtains
\begin{equation}
    a^r=\frac{\mu A'}{2A}.
\end{equation}
The acceleration measured locally by the static observer is the proper magnitude $a=(a_\mu a^\mu)^{1/2}$. Since $g_{rr}=1/\mu$, this gives
\begin{equation}
    a(r)
    =
    \frac{\sqrt{\mu}}{2}\frac{A'}{A}.
    \label{eq:proper_acceleration}
\end{equation}

 To define the corresponding surface-gravity scale with respect to the asymptotically normalized time coordinate, it is multiplied by the redshift factor $\sqrt{A(r)}$. We therefore define the generalized surface gravity as
\begin{equation}
    \kappa(r)
    \equiv
    \sqrt{A(r)}\,a(r).
    \label{eq:generalized_kappa_def}
\end{equation}
Using $A=e^{-2\delta}\mu$, this becomes
\begin{equation}
    \kappa(r)
    =
    e^{-\delta}
    \left(
    \frac{\mu'}{2}
    -\mu\delta'
    \right).
    \label{eq:generalized_kappa}
\end{equation}
Thus, $\kappa(r)$ represents the redshifted acceleration required to maintain a static observer at radius $r$, and extends the familiar surface-gravity construction away from the event horizon.

Using the Einstein equations \eqref{eq:mup} and \eqref{eq:deltap}, one finds
\begin{equation}
 \frac{\mu'}{2}-\mu\delta'
 =\frac{n-3}{2r}(1-\mu)+\frac{8\pi r}{n-2}p_r.
 \label{eq:kappaintermediate}
\end{equation}
Eliminating $p_r$ by means of Eq.~\eqref{eq:Rdef} gives a dimension-independent structural identity,
\begin{equation}
 \kappa(r)=\frac{\ee^{-\delta}}{r}
 \left(\mu-\frac{\mcR}{2}\right).
 \label{eq:kappaR}
\end{equation}
Thus, at the photon sphere,
\begin{equation}
 \kappa_\gamma=\frac{\ee^{-\delta_\gamma}\mu_\gamma}{r_\gamma}.
 \label{eq:kappagamma}
\end{equation}
It is worth stressing that Eq.~\eqref{eq:kappagamma} can also be obtained without using the field equations: Eq.~\eqref{eq:Aphoton} inserted into Eq.~\eqref{eq:proper_acceleration} immediately gives $a_\gamma=\sqrt{\mu_\gamma}/r_\gamma$, and multiplication by $\sqrt{A_\gamma}$ yields Eq.~\eqref{eq:kappagamma}.

Using Eq.~\eqref{eq:kappagamma}, the geometric bound \eqref{eq:rawbound} becomes
\begin{equation}
 \lambda\le\sqrt{n-3}\,
 \frac{\kappa_\gamma}{\sqrt{\mu_\gamma}}.
 \label{eq:kappabound}
\end{equation}

Since the radial null energy condition $\rho+p_r\ge0$ holds outside the photon sphere, Eq.~\eqref{eq:deltap} implies $\delta'\le0$.  Together with $\delta(\infty)=0$, this gives $\delta(r)\ge0$ at finite radius and hence $\ee^{-\delta_\gamma}\le1$.  Combining Eqs.~\eqref{eq:kappagamma} and \eqref{eq:rawbound} then yields
\begin{equation}
 \lambda^2\le(n-3)\frac{\kappa_\gamma}{r_\gamma}.
 \label{eq:kappabound2}
\end{equation}
This form relates the instability rate directly to the generalized surface gravity and the radius of the circular null orbit.

\section{Further consequences and observable forms}
\label{sec:applications}

\subsection{Shadow radius and orbital frequency}

The bound obtained above is expressed in terms of quantities evaluated locally at the circular null orbit. It is useful to recast this result in terms of quantities that have a more direct connection with the optical properties of the black hole. In particular, the critical impact parameter determines the apparent shadow radius for a distant observer, while the orbital frequency characterizes the angular motion of the unstable null orbit. We therefore express the Lyapunov bound in terms of these two quantities. The critical impact parameter associated with the circular null orbit is
\begin{equation}
 b_\gamma=\frac{L}{E}=\frac{r_\gamma}{\sqrt{A_\gamma}}
 =\frac{r_\gamma\ee^{\delta_\gamma}}{\sqrt{\mu_\gamma}}.
 \label{eq:bcrit}
\end{equation}
For a static, spherically symmetric, asymptotically flat black hole with the relevant unstable orbit setting the capture threshold, this is the shadow radius $R_{\rm sh}$ seen by an observer at infinity.  The coordinate angular velocity is
\begin{equation}
 \Omega_\gamma=\frac{\dot\phi}{\dot t}
 =\frac{\sqrt{A_\gamma}}{r_\gamma}
 =\frac{1}{R_{\rm sh}}.
 \label{eq:Omega}
\end{equation}
Equation \eqref{eq:rawboundsqrt} therefore becomes
\begin{equation}
 \lambda R_{\rm sh}\le\sqrt{n-3},
 \label{eq:shadowbound}
\end{equation}
 or equivalently
\begin{equation}
 \frac{\lambda}{\Omega_\gamma}\le\sqrt{n-3}.
 \label{eq:OmegaBound}
\end{equation}
The dimensionless combination $\lambda R_{\rm sh}$ cleanly measures the radial instability per optical timescale. The inequality therefore places a dimension-dependent upper limit on the 
instability rate measured relative to the optical scale of the black hole.

\subsection{Local acceleration and Unruh scale}

The generalized surface gravity introduced in the previous section is directly related to 
the proper acceleration required to maintain a static observer at a fixed radial position. It is therefore natural to ask whether the Lyapunov bound can be expressed directly in terms of this locally measured acceleration. Such a form provides a simple comparison between the instability timescale of the circular null orbit and the acceleration scale experienced by a static observer at the same radius. Moreover, through the Unruh effect, the same acceleration defines a local temperature scale, allowing the bound to be given an equivalent thermal interpretation.\\

From Eq.~\eqref{eq:proper_acceleration} and the photon-sphere condition \eqref{eq:Aphoton}, the proper acceleration of a static observer at the photon sphere is
\begin{equation}
 a_\gamma=\frac{\sqrt{\mu_\gamma}}{r_\gamma}.
 \label{eq:agamma}
\end{equation}
The bound \eqref{eq:rawboundsqrt} can be written
\begin{equation}
 \lambda\le\sqrt{n-3}\,\ee^{-\delta_\gamma}a_\gamma.
 \label{eq:accbound0}
\end{equation}
Under the radial NEC, $\ee^{-\delta_\gamma}\le1$, and therefore
\begin{equation}
 \lambda\le\sqrt{n-3}\,a_\gamma.
 \label{eq:accbound}
\end{equation}
Associating to the accelerated static observer the local Unruh temperature $T_U=a_\gamma/(2\pi)$ gives
\begin{equation}
 \lambda\le2\pi\sqrt{n-3}\,T_U.
 \label{eq:UnruhBound}
\end{equation}

The factor $\sqrt{n-3}$ determines how the maximal instability rate, measured relative to the local acceleration or Unruh temperature scale, changes with the dimensionality of spacetime.

\subsection{Critical exponent}

Another useful characterization of the instability of a circular null orbit is provided by the dimensionless critical exponent introduced by Cardoso \textit{et al.}~\cite{Cardoso2009}. It compares the orbital timescale with the instability timescale and therefore measures how rapidly nearby null trajectories diverge relative to the orbital motion. The critical exponent is defined as
\begin{equation}
    \gamma_c
    \equiv
    \frac{\Omega_\gamma}{2\pi\lambda}.
    \label{eq:critical_exponent}
\end{equation}
Using the bound obtained in Eq.~\eqref{eq:OmegaBound}, we immediately 
find
\begin{equation}
    \gamma_c
    \geq
    \frac{1}{2\pi\sqrt{n-3}}.
    \label{eq:critical_bound}
\end{equation}
Thus, the upper bound on the Lyapunov exponent translates directly into a dimension-dependent lower bound on the critical exponent. The Schwarzschild-Tangherlini geometry saturates this inequality, reproducing  the corresponding exact value obtained in Ref.~\cite{Cardoso2009}.

\subsection{Eikonal quasinormal modes}

The instability of circular null orbits is closely connected to the quasinormal ringing of black holes. In the eikonal limit, the real part of the quasinormal-mode frequency is governed by the orbital frequency of the unstable null orbit, while its imaginary part, which determines  the damping rate, is controlled by the corresponding Lyapunov exponent. The bound obtained above can therefore be translated into a constraint  on the damping of eikonal quasinormal modes.\\

For minimally coupled perturbations in a regime where the standard eikonal geodesic/QNM correspondence holds, the quasinormal frequencies take the form \cite{Cardoso2009}
\begin{equation}
 \omega_{\ell q}\simeq \ell\Omega_\gamma
 -i\left(q+\frac12\right)\lambda,
 \qquad \ell\gg1,
 \label{eq:eikonalQNM}
\end{equation}
where $q=0,1,2,\ldots$ is the overtone number.  Combining Eq.~\eqref{eq:eikonalQNM} with Eq.~\eqref{eq:shadowbound} yields
\begin{equation}
 |\operatorname{Im}\omega_{\ell q}|\,R_{\rm sh}
 \le\left(q+\frac12\right)\sqrt{n-3}.
 \label{eq:ImQNMbound}
\end{equation}
Since $\operatorname{Re}\omega_{\ell q}\simeq\ell/R_{\rm sh}$,
\begin{equation}
 \frac{|\operatorname{Im}\omega_{\ell q}|}
 {\operatorname{Re}\omega_{\ell q}}
 \lesssim\frac{q+\tfrac12}{\ell}\sqrt{n-3}.
 \label{eq:QNMratio}
\end{equation}
Equivalently, the ringdown quality factor $Q\equiv\operatorname{Re}\omega/[2|\operatorname{Im}\omega|]$ obeys the eikonal lower bound
\begin{equation}
 Q\gtrsim\frac{\ell}{2\sqrt{n-3}\,(q+\tfrac12)}.
 \label{eq:Qbound}
\end{equation}
These results should be interpreted with the usual qualification. The null-geodesic/QNM correspondence applies to test fields in the eikonal limit under the standard assumptions on the effective potential, but it need not hold for gravitational perturbations in higher-curvature theories. In particular, Konoplya and Stuchl\'ik showed that the correspondence is violated for gravitational perturbations of Einstein-Lovelock black holes~\cite{KonoplyaStuchlik2017}. We emphasize that the Lyapunov bounds derived in Eqs.~\eqref{eq:rawbound}-\eqref{eq:OmegaBound} concern the instability of circular null geodesics in $n$-dimensional Einstein gravity and are independent of the QNM correspondence. The latter is required only when these geodesic bounds are translated into the quasinormal-mode constraints in Eqs.~\eqref{eq:ImQNMbound}-\eqref{eq:Qbound}.
\section{Consistency checks and stronger corollaries}
\label{sec:checks}

\subsection{Four-dimensional limit}

As a consistency check, we now consider the four-dimensional limit of the preceding results. Setting $n=4$ in Eq.~\eqref{eq:RprimeExact} gives
\begin{equation}
    \mcR'(r_\gamma)
    =
    \frac{2}{r_\gamma}
    \left[
    1-8\pi r_\gamma^2
    (\rho_\gamma+p_{t\gamma})
    \right],
    \label{eq:Rprime4D}
\end{equation}
which coincides with the local photon-sphere relation employed in the four-dimensional analysis of Ref.~\cite{GalloMadler2025}. Consequently, the dimension-dependent Lyapunov bounds obtained above reduce to
\begin{equation}
    \lambda\leq\frac{\kappa_\gamma}{\sqrt{\mu_\gamma}},
    \qquad
    \lambda^2\leq\frac{\kappa_\gamma}{r_\gamma},
    \qquad
    \lambda R_{\rm sh}\leq1.
    \label{eq:4Dbounds}
\end{equation}
The corresponding bounds involving the orbital frequency, local acceleration, critical exponent, and eikonal quasinormal modes likewise follow by setting $n=4$ in the respective expressions derived above. Thus, the four-dimensional results are recovered consistently, while the factor $\sqrt{n-3}$ appearing in the general expressions captures the dimensional dependence of the higher-dimensional bounds.

\subsection{Schwarzschild-Tangherlini saturation}

For the $n$-dimensional Schwarzschild-Tangherlini vacuum solution \cite{Tangherlini1963},
\begin{equation}
 \delta=0,
 \qquad
 \mu(r)=1-\frac{2M}{r^{n-3}},
 \label{eq:Tangherlini}
\end{equation}
with $\rho=p_r=p_t=0$.  Equation \eqref{eq:RprimeExact} therefore saturates Eq.~\eqref{eq:RprimeBound}.  The photon-sphere condition gives
\begin{equation}
 r_\gamma^{n-3}=(n-1)M,
 \qquad
 \mu_\gamma=\frac{n-3}{n-1}.
 \label{eq:TangherliniPhoton}
\end{equation}
Consequently,
\begin{equation}
 \lambda=\sqrt{n-3}\,\Omega_\gamma,
 \label{eq:TangherliniSaturation}
\end{equation}
which agrees with the exact higher-dimensional result of Cardoso \emph{et al.} \cite{Cardoso2009}.  The vacuum solution therefore saturates Eqs.~\eqref{eq:rawbound}, \eqref{eq:shadowbound}, \eqref{eq:OmegaBound}, and \eqref{eq:critical_bound}.  This sharpness is important: the factor $\sqrt{n-3}$ is not an artifact of a loose estimate but is fixed by the vacuum solution itself.\\

It is worth comparing this result with the higher-dimensional analysis 
of Bianchi, Grillo and Morales~\cite{Bianchi:2020des}. For spherically symmetric charged black holes with $g_{tt}g_{rr}=-1$, they obtained an upper bound on the Lyapunov exponent in terms of the critical impact parameter $b_c$. After accounting for the different normalization of the Lyapunov exponent adopted in Ref.~\cite{Bianchi:2020des}, their result can be written as
\begin{equation}
    \lambda b_c\leq\sqrt{n-3}.
\end{equation}
For asymptotically flat spherically symmetric spacetimes, $b_c=R_{\rm sh}$, and hence this agrees with the shadow bound expressed in Eq. \ref{eq:shadowbound}. The present derivation, however, does not rely on a particular charged black-hole solution or on the restriction $g_{tt}g_{rr}=-1$. Instead, the same dimension-dependent bound follows for a general static, spherically symmetric solution of the Einstein equations satisfying the corresponding energy condition. The earlier result is therefore recovered as a particular case of the more general geometric bound derived here.

\subsection{Bounds using stronger matter assumptions}

The core Lyapunov bound requires only the tangential NEC at the circular null orbit.  Stronger geometric bounds follow if one adopts the assumptions used by Song, Fu, and Cen \cite{SongFuCen2026}.  Under their weak/dominant-type inequalities together with a non-positive trace of the energy-momentum tensor, one obtains
\begin{equation}
 p_r(r_\gamma)\le0,
 \qquad
 \mu_\gamma\le\frac{n-3}{n-1}.
 \label{eq:muSong}
\end{equation}
Combining the second inequality with Eq.~\eqref{eq:rawboundsqrt} and $\ee^{-\delta_\gamma}\le1$ gives the radius-based bound
\begin{equation}
 \lambda\le\frac{n-3}{\sqrt{n-1}}\frac{1}{r_\gamma}.
 \label{eq:radiustight}
\end{equation}
If, in addition, the monotonicity condition used for the lower photon-sphere-radius theorem of Ref.~\cite{SongFuCen2026} is imposed, then
\begin{equation}
 r_\gamma\ge
 \left(\frac{n-1}{2}\right)^{1/(n-3)}r_H,
 \label{eq:lowerPhotonSong}
\end{equation}
leading to
\begin{equation}
 \lambda r_H\le
 \frac{n-3}{\sqrt{n-1}}
 \left(\frac{2}{n-1}\right)^{1/(n-3)}.
 \label{eq:horizoncorollary}
\end{equation}
Unlike the principal result \eqref{eq:kappabound}, Eq.~\eqref{eq:horizoncorollary} depends on additional global matter assumptions and should therefore be regarded as a stronger but less universal corollary.
\section{Summary and Discussion}
\label{sec:conclusions}

In this work, we have investigated the instability of circular null geodesics in static, spherically symmetric, asymptotically flat black-hole spacetimes in $n$-dimensional Einstein gravity. For a general anisotropic matter distribution, we have derived dimension-dependent upper bounds on the Lyapunov exponent governing the radial instability of these orbits. The analysis shows how the allowed instability rate is constrained by the spacetime dimension, the local geometry at the photon sphere, and the energy conditions satisfied by the matter fields.\\

The key observation is that, although the higher-dimensional photon-sphere characteristic function $\mcR(r)$ depends explicitly on the radial pressure, this dependence disappears from its derivative when stress-energy conservation is evaluated at the photon sphere. As shown in Eq.~\eqref{eq:RprimeExact}, $\mcR'(r_\gamma)$ depends on the matter sector only through the combination $\rho_\gamma+p_{t\gamma}$. Consequently, the tangential null energy condition is sufficient to constrain the instability of the circular null orbit, without requiring an independent assumption on the sign of the radial pressure at $r=r_\gamma$. The resulting principal bound can be expressed in terms of the generalized surface gravity at the photon sphere as
\begin{equation}
    \lambda
    \leq
    \sqrt{n-3}\,
    \frac{\kappa_\gamma}{\sqrt{\mu_\gamma}}.
    \label{eq:summarymainbound}
\end{equation}
Additional forms of this inequality were obtained in terms of the generalized surface gravity and photon-sphere radius, the proper acceleration of a static observer, the shadow radius, and the angular frequency of the circular null orbit. In particular, the dimensionless bounds given in Eqs.~\eqref{eq:shadowbound} and \eqref{eq:OmegaBound} provide a direct relation between the instability timescale and the optical properties of the circular null orbit. Under the standard eikonal geodesic-quasinormal-mode correspondence, these results also lead to constraints on the damping rate and quality factor of the corresponding quasinormal modes.

All the principal bounds reduce to the corresponding four-dimensional results of Gallo and M\"adler~\cite{GalloMadler2025} when $n=4$. Moreover, the Schwarzschild-Tangherlini solution saturates the dimensionless Lyapunov bound, reproducing the known relation between the Lyapunov exponent and orbital frequency obtained by Cardoso \emph{et al.}~\cite{Cardoso2009}. The appearance of the factor $\sqrt{n-3}$ is therefore not merely a consequence of the inequalities used in the derivation, but characterizes the dimensional dependence of the limiting vacuum solution.

An interesting feature of the result is that increasing the spacetime dimension weakens the normalized upper bound on the instability of the circular null orbit. In this sense, the dimensional dependence enters through a simple multiplicative factor, while the influence of matter at the photon sphere is governed locally by the tangential null energy condition. The bound therefore provides a useful benchmark for comparing photon-sphere instability in higher-dimensional black-hole geometries.

The connection with observable quantities should, however, be interpreted with some caution. Current photon-ring and gravitational-wave observations concern astrophysical black holes that are consistent with four-dimensional general relativity and provide no evidence for extra spacetime dimensions. Nevertheless, the relation of the Lyapunov exponent to photon-ring structure and, under appropriate conditions, to quasinormal-mode damping makes such bounds useful for identifying theoretical consistency relations among strong-field quantities. In this respect, the higher-dimensional results obtained here may serve as a reference point for investigating how these relations are modified in gravitational theories beyond four-dimensional Einstein gravity.

An important direction is to extend the analysis beyond Einstein gravity. Einstein-Gauss-Bonnet and, more generally, Lovelock theories are particularly relevant in higher dimensions. The modified field equations change the relation between the matter variables, the metric functions, and the photon-sphere characteristic function. Although bounds on the photon-sphere radius have already been studied in these theories, it remains to be seen whether a corresponding dimension-dependent bound on the Lyapunov exponent can be established.

A more challenging extension concerns rotating higher-dimensional black 
holes. Once spherical symmetry is lost, null trapping is generally described by a photon region rather than a single photon sphere, making the construction of a universal instability bound considerably more involved. Investigating these extensions would help determine whether the bounds derived here reflect a more general property of black-hole spacetimes or are specific to the static, spherically symmetric setting considered in this work.

\begin{acknowledgments}
This work is funded by Institutional Faculty Research Seed Grant Scheme (IFRSG-2026), TKM College Trust. The authors thank Md.~Sabir Ali and C.~L.~Ahmed Rizwan for useful discussions.
\end{acknowledgments}

\bibliographystyle{apsrev4-2}
\bibliography{lyapunov_bounds_nd}

\appendix

\section{Derivation of the radial pressure equation}
\label{app:pressure_conservation}

In this section, we provide a detailed derivation that leads to Eq. \ref{eq:conservationA} used in Sec.~IV. The result follows directly from the conservation of the energy-momentum tensor and is useful for relating the matter variables to the geometry at the circular null orbit. For convenience, we write the metric in the form
\begin{equation}
    ds^2
    =
    -A(r)dt^2
    +\frac{dr^2}{\mu(r)}
    +r^2d\Omega_{n-2}^2,
    \label{eq:app_metric}
\end{equation}
where
\begin{equation}
    A(r)=e^{-2\delta(r)}\mu(r).
    \label{eq:app_A}
\end{equation}
The anisotropic energy-momentum tensor introduced in the main text is
\begin{equation}
    T^\mu{}_\nu
    =
    \operatorname{diag}
    \left(
    -\rho,p_r,p_t,\ldots,p_t
    \right).
    \label{eq:app_stress}
\end{equation}
The radial pressure is determined by the radial component of the
conservation equation,
\begin{equation}
    \nabla_\mu T^\mu{}_r=0.
    \label{eq:app_conservation}
\end{equation}

For a mixed rank-two tensor, the covariant derivative can be expanded as
\begin{equation}
    \nabla_\mu T^\mu{}_\nu
    =
    \partial_\mu T^\mu{}_\nu
    +\Gamma^\mu{}_{\mu\lambda}T^\lambda{}_\nu
    -\Gamma^\lambda{}_{\mu\nu}T^\mu{}_\lambda.
    \label{eq:app_covariant}
\end{equation}
Taking $\nu=r$, Eq.~\eqref{eq:app_covariant} becomes
\begin{equation}
\begin{split}
    \nabla_\mu T^\mu{}_r
    ={}&
    \partial_\mu T^\mu{}_r
    +\Gamma^\mu{}_{\mu\lambda}T^\lambda{}_r
    \\
    &-\Gamma^\lambda{}_{\mu r}T^\mu{}_\lambda.
\end{split}
\label{eq:app_radial_expansion}
\end{equation}
Since the energy-momentum tensor is diagonal, $T^\mu{}_r$ is nonzero
only for $\mu=r$. The first term in
Eq.~\eqref{eq:app_radial_expansion} therefore reduces to
\begin{equation}
    \partial_\mu T^\mu{}_r
    =
    \partial_r T^r{}_r
    =
    p_r'.
    \label{eq:app_first}
\end{equation}

The same diagonality-requirement considerably simplifies the second term. Since
$T^\lambda{}_r$ is nonzero only when $\lambda=r$, we have
\begin{equation}
    \Gamma^\mu{}_{\mu\lambda}T^\lambda{}_r
    =
    p_r\Gamma^\mu{}_{\mu r}.
    \label{eq:app_second}
\end{equation}
The sum over $\mu$ includes the temporal and radial coordinates together
with all $(n-2)$ angular coordinates. Hence,
\begin{equation}
\begin{split}
    \Gamma^\mu{}_{\mu\lambda}T^\lambda{}_r
    =
    p_r\Bigg(
    \Gamma^t{}_{tr}
    +\Gamma^r{}_{rr}
    +\sum_{i=1}^{n-2}
    \Gamma^{\theta_i}{}_{\theta_i r}
    \Bigg).
\end{split}
\label{eq:app_second_expanded}
\end{equation}

For the last term in Eq.~\eqref{eq:app_radial_expansion}, the diagonal
form of $T^\mu{}_\lambda$ requires $\lambda=\mu$. Using
$T^t{}_t=-\rho$, $T^r{}_r=p_r$, and
$T^{\theta_i}{}_{\theta_i}=p_t$, we find
\begin{equation}
\begin{split}
    -\Gamma^\lambda{}_{\mu r}T^\mu{}_\lambda
    ={}&
    \rho\,\Gamma^t{}_{tr}
    -p_r\Gamma^r{}_{rr}
    \\
    &-p_t\sum_{i=1}^{n-2}
    \Gamma^{\theta_i}{}_{\theta_i r}.
\end{split}
\label{eq:app_third}
\end{equation}

Combining Eqs.~\eqref{eq:app_first}, \eqref{eq:app_second_expanded}, and \eqref{eq:app_third}, the terms containing $\Gamma^r{}_{rr}$ cancel. The radial conservation equation then takes the simpler form
\begin{equation}
\begin{split}
    0={}&p_r'
    +(\rho+p_r)\Gamma^t{}_{tr}
    \\
    &+(p_r-p_t)
    \sum_{i=1}^{n-2}
    \Gamma^{\theta_i}{}_{\theta_i r}.
\end{split}
\label{eq:app_before_christoffel}
\end{equation}
Thus, only the temporal and angular connection coefficients are needed to obtain the desired relation.

For the temporal part of the metric, $g_{tt}=-A(r)$, and therefore
\begin{equation}
    \Gamma^t{}_{tr}
    =
    \frac{1}{2}g^{tt}\partial_r g_{tt}
    =
    \frac{A'}{2A}.
    \label{eq:app_christoffel_t}
\end{equation}
For the angular sector, each diagonal component of the metric contains the same overall factor $r^2$. Its radial derivative consequently gives
\begin{equation}
    \Gamma^{\theta_i}{}_{\theta_i r}
    =
    \frac{1}{r},
    \qquad i=1,\ldots,n-2.
    \label{eq:app_christoffel_ang}
\end{equation}
Since there are $(n-2)$ angular directions, their total contribution is
\begin{equation}
    \sum_{i=1}^{n-2}
    \Gamma^{\theta_i}{}_{\theta_i r}
    =
    \frac{n-2}{r}.
    \label{eq:app_angular_sum}
\end{equation}
The factor $(n-2)$ appearing in the pressure equation therefore has a simple geometrical origin: it counts the angular directions of the $n$-dimensional spherically symmetric spacetime.

Substitution of Eqs.~\eqref{eq:app_christoffel_t} and \eqref{eq:app_angular_sum} into Eq.~\eqref{eq:app_before_christoffel} gives
\begin{equation}
    p_r'
    +\frac{A'}{2A}(\rho+p_r)
    +\frac{n-2}{r}(p_r-p_t)
    =0.
    \label{eq:app_pressure_intermediate}
\end{equation}
Finally, solving for the radial derivative of the pressure yields
\begin{equation}
    p_r'
    =
    -\frac{A'}{2A}(\rho+p_r)
    +\frac{n-2}{r}(p_t-p_r).
    \label{eq:app_pressure_final}
\end{equation}

Equation~\eqref{eq:app_pressure_final} is the $n$-dimensional anisotropic pressure-conservation equation used in Sec.~IV. At
the circular null orbit, the photon-sphere condition $A'(r_\gamma)/A(r_\gamma)=2/r_\gamma$ further simplifies this relation, which is the form required in deriving the upper bound on the Lyapunov exponent.

\end{document}